\documentstyle[epsf,prd,twocolumn,aps,floats]{revtex} 
\tightenlines
\begin{document}

\draft
\topmargin = -0.6cm
\topmargin = -2.0cm
\overfullrule 0pt
\twocolumn[\hsize\textwidth\columnwidth\hsize\csname
@twocolumnfalse\endcsname

\title{ 
\vglue -0.5cm
\hfill{\small IFT-P.053/2001} \\
\vglue 0.5cm
 Neutrino-neutrino and neutrino-matter helicity flip interactions }

\author{\bf Vicente Pleitez
}
\address{
Instituto de F\'\i sica Te\'orica\\
Universidade Estadual Paulista\\
Rua Pamplona, 145\\ 
01405-900-- S\~ao Paulo, SP\\
Brazil}
\date{\today}
\maketitle
\vspace{.5cm}

\hfuzz=25pt

\begin{abstract}
Taking into account that neutrinos are massive particles and that they are
produced mainly as states of negative helicity, we show that the neutral and
charged current interactions change these neutrinos into transversally polarized
states. 
This implies a considerable reduction of the neutrino flux when
propagating through ordinary matter (electrons, protons and neutrons).
The same happens when neutrinos propagate through the sea of relic neutrinos if
these neutrinos are degenerate fermions. However, in this case the change of
helicity depends on the value of the neutrino--anti-neutrino asymmetry.
\end{abstract}


\vskip2pc]

\narrowtext







The recent solar~\cite{nussol,sno} and atmospheric~\cite{nusat} neutrino data
are consistent with three neutrinos with masses $m_1=0$, $m_2=2.3\times10^{-3}$
eV and $m_3=5\times10^{-2}$ eV. 
In this paper we show that if neutrinos are massive particles, the neutral and
charged effective interactions due to $Z^0$ and $W^\pm$ exchange will induce an
helicity flip on the neutrinos of negative helicity.
These effects are complementary to those discovered, some years ago, by  
Stodoslky~\cite{ls} and Wolfenstein~\cite{lw1}. 

Let us first consider neutrinos coming
from astrophysical sources, here denoted by $\nu_A$. They interact with the
relic cosmological neutrinos, denoted by $\nu_r$, through the effective
hamiltonian (induced by the neutral current coupled to $Z^0$) 
\begin{equation}
H_Z=\frac{G_F}{\sqrt2 c^2_W} \bar{\psi}_{\nu_A}\gamma_\mu(1-\gamma_5)
\psi_{\nu_A} \bar{\psi}_{\nu_r}\gamma_\mu(1-\gamma_5)\psi_{\nu_r}.
\label{e1}
\end{equation}
We will also assume that neutrinos are Dirac fermions. 

If the relic neutrino is massive we can use its rest frame. Since
$T_\nu=195\,{\rm K}\approx1.68\times10^{-4}$ eV, neutrinos with masses
$m_{2,3}>T_\nu$ are non-relativistic at present. However, since our goal is
to point out new effects we are only interested in the order of magnitude, as an
illustration of the phenomena, we will assume for the sake of simplicity that
the relic neutrino, $\nu_r$, is the lightest neutrino, i.e., that with $m_1=0$,
and that the astrophysical neutrinos $\nu_A$ are massive
neutrinos, i.e., $\nu_2$ or $\nu_3$. In the rest frame of the ``neutrino sea'',
in which the relic neutrinos are isotropically distributed, the term 
$\bar{\psi}_{\nu_r}\gamma_\mu(1-\gamma_5)\psi_{\nu_r}$ is the neutrino
number density (in fact, $n_\nu-n_{\bar\nu}$) and for the relic neutrinos
$1-\gamma_5=2$.
In the same frame the astrophysical neutrino has a velocity $\vec{v}$ and spin
$\vec{\sigma}$; thus the corresponding bilinear is proportional to
$1+\vec{\sigma}\cdot\vec{v}$. On the other hand, in the frame of the massive
neutrino the relic neutrinos have a current density
$\vec{J}_\nu=2n_\nu\vec{v}\gamma$, where
$\gamma=(1-v^2)^{-1/2}=E_{\nu_A}/m_{\nu_A}$ (we are using $\hbar=c=1$). Since
the helicity is an eigenstate of propagation, we can write the evolution 
equation in the source basis denoted by primed fields 
\begin{equation}
i\frac{d}{d\tau} \left( 
\begin{array}{c}
\nu^\prime_+ \\ \nu^\prime_-
\end{array}
\right)=\left[\left(\begin{array}{cc}
E &0\\
0&E\end{array}\right)+\frac{2G_F n_\nu\gamma}{\sqrt2 c^2_W}
\,\vec{\sigma}\cdot\vec{v}\right]
\left( 
\begin{array}{c}
\nu^\prime_+ \\ \nu^\prime_-
\end{array}
\right),
\label{e2}
\end{equation}
where $\tau$ is the proper time of the astrophysical neutrino. 
Neglecting an overall phase factor $E$ (we are assuming $E'_+=E'_-\equiv E$,
otherwise the global phase is $(E'_++E'_-)/2$) we can re-write Eq.~(\ref{e2}) as
\begin{equation}
i\frac{d}{d\tau} \left( 
\begin{array}{c}
\nu^\prime_+ \\ \nu^\prime_-
\end{array}
\right)=\frac{2G_F n_\nu\gamma v_z}{\sqrt2 c^2_W}\left(\begin{array}{cc}
1 & \tan\theta e^{-i\phi} \\
\tan\theta e^{i\phi} &-1
\end{array}\right)
\left( 
\begin{array}{c}
\nu^\prime_+ \\ \nu^\prime_-
\end{array}
\right),
\label{e3}
\end{equation}
where $\tan\theta=2v_\bot/v_z$  with
$v_\bot=(v^2_x+v^2_y)^{1/2}$ and $\phi=\arctan (v_y/v_x)$; notice that
$0\leq \theta<\pi$ and $0\leq \phi<2\pi$. 
The eigenvalues of the matrix in Eq.~(\ref{e3}) are given by
\begin{equation}
E_\pm=\pm\frac{2G_F n_\nu\,v\gamma}{\sqrt2 c^2_W },
\label{e4}
\end{equation}
and 
\begin{equation}
\left(\begin{array}{c}
\nu_+\\\nu_-\end{array}\right)=
\left(\begin{array}{cc}\cos\frac{\theta}{2}\,e^{-i\frac{\phi}{2}} &
\sin\frac{\theta}{2}\,e^{i\frac{\phi}{2}}\\ 
-\sin\frac{\theta}{2}\,e^{-i\frac{\phi}{2}} &
\cos\frac{\theta}{2}\,e^{i\frac{\phi}{2}}\end{array} 
\right)\left(\begin{array}{c}
\nu^\prime_+\\ \nu^\prime_-\end{array}\right),
\label{e5}
\end{equation}
denote the basis in which the hamiltonian is diagonal. 

Astrophysical neutrinos are produced, as usual, via the standard electroweak
processes, with are mainly of negative helicity $\nu^\prime_-$, however they
become, because of the interaction with the sea, a linear superposition of both
helicity states
\begin{equation}
\vert \nu^\prime_-\rangle=e^{-i\phi}\left(e^{-iE_+\tau}
\sin\frac{\theta}{2}\,\vert \nu_+\rangle + 
e^{-iE_-\tau}\cos\frac{\theta}{2}\,\vert \nu_-\rangle\right),
\label{e6}
\end{equation}
and we can calculate for the massive neutrinos the survival probability (they
remain in the same helicity state):
\begin{equation}
P(\nu^\prime_-\to\nu^\prime_-)=1-\sin^2\theta\sin^2\left(\frac{4G_F n_\nu
}{\sqrt2 c^2_W}d \right),
\label{pro1}
\end{equation}
and the transition probability (when there is a helicity flip)
\begin{equation}
P(\nu^\prime_-\to\nu^\prime_+)=1-P(\nu^\prime_-\to\nu^\prime_-)
\label{pro2}
\end{equation}
where $(E_+-E_-)t/\gamma$ becomes $(4G_F n_\nu d)/\sqrt{2}c^2_W $, in the rest
frame of the relic neutrinos $\tau=t\gamma^{-1}$, and we have defined
$d=vt$ as the distance traveled through the relic neutrino sea. 

If both $\nu$ and $\bar{\nu}$ are present $n_\nu$ means
$n_\nu-n_{\bar{\nu}}$. However we can also assume that there is an asymmetry
i.e., neutrinos form a degenerate Fermi gas, $n_\nu\not=0$ and
$n_{\bar{\nu}}\approx0$ (or vice versa).  
The oscillation length defined from Eq.~(\ref{pro1}) is
\begin{equation}
l\equiv \frac{\pi}{2}\, \frac{\sqrt{2} c^2_W}{4G_Fn_\nu}, 
\label{oscil}
\end{equation}
and assuming $n_\nu\approx1700 \,{\rm
cm}^{-3}(1.3\times10^{-11}\,{\rm eV}^3)$~\cite{frances} we see that
$l=5.28\times10^{9}$ ly. Which is of the order of the radius of the observed
universe. 
Notice, however,  that for $d=0.5l$ we have from Eq.~(\ref{pro1}) that
$P(\nu^\prime_-\to\nu^\prime_-)=0.5$ for $\theta=\pi/2$. Moreover, if the relic
neutrinos are the massive ones they can be clustered in local galactic halo
and it is possible to have $n_\nu=(10^7-10^9)\,{\rm cm}^{-3}$~\cite{mele}. Using,
for instance, $n_\nu=10^7$ the oscillation length is $l=1.05\times10^6$ ly,
i.e., this clustering, if it does exist, reduce the oscillation length at least
in some regions. (But in this case, relic  neutrinos are non-relativistic and
the approach of Ref.~\cite{dudas} has to be used.) If the value of $l$ can be
lower than the size of the universe we have to take also the average over the
distance, the average survival probability is  
\begin{equation}
P\equiv \langle P(\nu^\prime_-\to\nu^\prime_-)\rangle
=1-\frac{1}{2}\sin^2\theta. 
\label{proave}
\end{equation}
The average probability in $\theta$ may also be needed in some cases: 
\begin{equation}
P_\theta\equiv\langle P(\nu^\prime_-\to\nu^\prime_-)\rangle_\theta =0.75.
\label{proave2}
\end{equation}

As an effect due to the potential it depends only on the number density of
particles in the medium and on the distance traveled by the propagating
particle. The effect depends also on a non-zero, but otherwise arbitrary,
neutrino mass.
The same happens also for charged particles in cosmic rays, say
electrons, however in this case both helicity states
feel the electromagnetic interactions with the same strength so the effect is,
in practice, usefulness. Neutrinos with positive helicity will have a similar
effect, but since these neutrinos are produced by exotic unknown
interactions their fluxes are reduced with respect to the flux of neutrinos
produced in the usual electroweak processes.

The case when neutrinos propagate through ordinary matter is more interesting.
We will consider only the charged current interaction (after a Fierz
transformation) since here we are only showing the
main features of the effect. The effective hamiltonian is
\begin{equation}
H_W=\frac{G_F}{\sqrt2} \bar{\psi}_{e}\gamma_\mu(1-\gamma_5)
\psi_{e} \bar{\psi}_{\nu}\gamma_\mu(1-\gamma_5)\psi_{\nu},
\label{e8}
\end{equation}
$\nu$ is one of the massive neutrinos and we are neglecting mixing angles.
We have with this interaction a flip of the electron spin due to massless
neutrinos (Stodolsky effect); or the effect in the propagation 
through matter of massless or massive neutrinos (Wolfenstein effect); and also
the spin flip of a massive neutrino due to the matter density, this is the
effect that we will consider below.
Similarly to Eq.~(\ref{pro1}) but without the $2/c^2_W$ factor 
and $n_e$ being now the electron number density in the medium, say the Sun, the
Earth, the atmosphere or even a detector. The oscillation length is now given
by 
\begin{equation}
l\equiv\frac{\pi}{2}\,\frac{1}{\sqrt{2}G_Fn_e} ,
\label{lele}
\end{equation}
since in this case $\Delta E=\sqrt2 G_F n_ev\gamma$, and $v$ is now the neutrino
velocity in the rest frame of the electrons. In the neutrino rest frame the
electron current density is $\vec{J}_e=n_e\vec{v}\gamma$. Assuming a typical
value $\rho=1\,{\rm gr/cm^3}$ (or $n_e\approx 8.45\times10^{12}\,{\rm eV}^3$) we
see that the oscillation length is  in this case $l\approx2.2$ km. 
In the solar interior $l$ is two order of magnitude smaller that this value
while in the atmosphere it is two or three order of magnitude larger.
Hence in solar and atmospheric neutrino experiments only the averaged
probability transition in Eq.~(\ref{proave}) may be considered. A similar effect
occur with protons through both neutral and charged currents and with
neutrons through the neutral currents only.

We can interpret these effects, induced by the interactions in
Eqs.~(\ref{e1}) or (\ref{e8}), as follows. The usual neutrino oscillations which
are driven by mass square differences can be understood as a transition 
between a left-handed neutrino of flavor f to a left-handed neutrino of flavor
f$^\prime$ (f may be equal or different from f$^\prime$), via a negative
helicity neutrino mass eigenstate $\nu'_-$
\begin{equation}
\nu_{\rm f L}\to \nu'_{\rm -}\to\nu_{\rm f^\prime L}.
\label{inter}
\end{equation}
In the standard model, flavor neutrinos are produced as
left-handed states, the intermediate mass eigenstates are also mainly of
negative helicity because of $\nu_{fL}=\nu'_{-}+O(m/E_\nu)$. Since neutrino
masses are small and the respective energies, for the usual experiments, are
large we see that $\nu_{fL}\approx \nu'_{-}$. However, in the presence of matter
(or relic neutrinos) there is an helicity flip $\nu'_{-}\to \nu'_{+}$ and this
transition  is independent of the neutrino mass, it depends only on the matter
(sea) density, but again $\nu'_{+}\approx \nu_{fR}+ O(m/E_\nu)$ the neutrino is
not detected since it is mainly an (sterile) right-handed neutrino. 

Hence, as an illustration, in the two flavor case and assuming $m_1=0$ and
$m_2\not=0$ (but otherwise arbitrary) we should write
\begin{equation}
\left( \begin{array}{c}
\nu_e \\ \nu_\mu\end{array}\right)=\left( \begin{array}{cc}
\cos\theta_f & P\sin\theta_f \\
-\sin\theta_f & P\cos\theta_f\end{array}\right)
\left( \begin{array}{c}
\nu_1 \\ \nu_2\end{array}\right),
\label{ufa}
\end{equation} 
where $\theta_f$ is a Cabibbo-like mixing angle. For geochemical experiments
$P=P_\theta=0.75$ must be used. On the other hand, in directional, real time
experiments in which the $z$-axis is already defined as the azimuthal
direction, we can identify $\theta$ with the azimuthal angle and
Eq.~(\ref{proave}) must be used. Only vacuum neutrino oscillation experiments are
not sensitive to the effect induced by the hamiltonian given in Eq.~(\ref{e8}).
In this case the oscillation length is large enough (induced only by the relic
neutrinos) to be appreciable at the typical distances of this sort of
experiments. This is the case for the LSND experiment~\cite{lsnd}.

The considerations above are valid for Dirac neutrinos. In this case the
right-handed chiral states are sterile neutrinos, $\nu_R$. For Majorana
neutrinos $\nu_R$'s are not sterile but interact mainly with anti-leptons. In
this case in order to have a non-vanishing effect in the neutrino-neutrino
interactions, it is necessary to have a net left-handed chirality asymmetry, 
$n_{\nu_L}\not=0$, $n_{\nu_R}\approx0$. The Stodolsky effect with Majorana relic
neutrinos was considered also in Ref.~\cite{dudas}.

The mechanical forces exerted on macroscopic targets by relic neutrinos is
unfortunately very small~\cite{dudas,maiani}. However we have considered here
the effect on neutrinos propagating in a medium which can be the
relic neutrino sea or ordinary charged or neutral matter. The important thing,
besides a non-zero but arbitrary mass, is the
number density of the medium and the distance travel by the neutrinos. The
helicity flip does not depend on the real value of the neutrino masses.

There are also spin-spin interactions (i.e., the effect of a vector potential)
but this case needs a more careful treatment.
We would like to call the effects above considered ``GE$_F$AN--effects''.

\acknowledgments 
This work was supported by Funda\c{c}\~ao de Amparo \`a Pesquisa
do Estado de S\~ao Paulo (FAPESP), Conselho Nacional de 
Ci\^encia e Tecnologia (CNPq) and by Programa de Apoio a
N\'ucleos de Excel\^encia (PRONEX). The author would like to thank G. Matsas,
all the GE$_F$AN and IFT-UNESP collaborators for useful discussions.


\begin{references}
\bibitem{nussol}  
B. T. Cleveland {\it et al.} (Homestake Collaboration),
Astrophys. J. {\bf496},  505 (1998); K. S. Hirata {\it et al.} (Kamiokande 
Collaboration), Phys. Rev. Lett. {\bf77}, 1683 (1996); W. Hampel {\it et al.}
(GALLEX Collaboration), Phys. Lett. {\bf B477}, 127 (1999); 
J. N. Abdurashitov {\it at al.} (SAGE Collaboration), 
Phys. Rev. Lett. {\bf77}, 4708 (1996); Phys. Rev. C  {\bf60}, 055801 (1999).
\bibitem{sno} Q. R. Ahmad {\it et al.} (SNO Collaboration), nucl-ex/0106015.
\bibitem{nusat} Y. Fukuda {\it et al.} (SuperKamiokande Collaboration),
Phys. Rev. Lett. {\bf81}, 1562 (1998);
{\it ibid}, {\bf82}, 2644 (1999). 
\bibitem{ls} L. Stodolsky, Phys. Rev. Lett. {\bf34}, 110 (1975).
\bibitem{lw1} L. Wolfenstein, Phys. Rev. D {\bf17}, 2369 (1978).
\bibitem{frances} J. Lesgourgues and S. Pastor, Phys. Rev. D 
{\bf60}, 103521 (1999). 
\bibitem{mele} D. Fargion, B. Mele and A. Salis, Astrophys. J.
{\bf517}, 725 (1999). 
\bibitem{dudas} G. Dudas, G. Gelmini and S. Nussinov, hep-ph/0107027.
\bibitem{lsnd} C. Athanassopoulos {\it et al.} (LSND Collaboration),
Phys. Rev. Lett. {\bf77}, 3082 (1996); {\it ibid} {\bf81}, 1774 (1998).
\bibitem{maiani} N. Cabibbo and L. Maiani, Phys. Let. {\bf114B}, 115 (1982);
P. Langacker, J. P. Leville and J. Sheiman, Phys. Rev. D {\bf27}, 1228 (1983).
\end{references}
\end{document}